\documentclass[12pt,a4paper]{article}
\usepackage{amsmath,caption}
\usepackage[nofiglist, notablist]{endfloat}
 
\usepackage{longtable,threeparttable,booktabs,}
\DeclareDelayedFloatFlavour*{longtable}{table}

\usepackage[top=1.37in, bottom=1.37in, left=1.1in, right=1.1in]{geometry}

\usepackage[round]{natbib}
\usepackage{color,soul}

\DeclareCaptionStyle{italic}[justification=centering]{labelfont={bf},textfont={it},labelsep=colon}
\usepackage{graphicx,psfrag,epsf,textcomp}
\usepackage{enumerate, dsfont}
\usepackage{natbib}
\usepackage{url,xcolor} 
\usepackage{booktabs, subfig, bm, paralist,mathpazo,tikz,todonotes,longtable,microtype}

\usepackage[pdftex,colorlinks=true]{hyperref}
\definecolor{darkblue}{rgb}{0,0,.6}
\hypersetup{citecolor=darkblue,linkcolor=darkblue,urlcolor=darkblue}
\newcommand{\argmax}{\operatornamewithlimits{argmax}}

\newcommand{\blind}{0}

\graphicspath{{plots/}}

\newsavebox\CBox
\def\textBF#1{\sbox\CBox{#1}\resizebox{\wd\CBox}{\ht\CBox}{\textbf{#1}}}

\begin{document}

\def\spacingset#1{\renewcommand{\baselinestretch}{#1}\small\normalsize} \spacingset{1}

\if0\blind
{
  \title{\bf Model confidence sets and forecast combination: \hbox{An application to age-specific mortality}}
  \author{
    Han Lin Shang\footnote{Postal address: Research School of Finance, Actuarial Studies and Statistics, Level 4, Building 26C, Kingsley Street, Australian National University, Acton Canberra, ACT 2601, Australia; Telephone: +61(2) 612 50535; Fax: +61(2) 612 50087; Email: hanlin.shang@anu.edu.au}\\
    Research School of Finance, Actuarial Studies and Statistics \\
 Australian National University \\
\\
 Steven Haberman \\
 Cass Business School \\
 City, University of London\\
 }
  \maketitle
} \fi

\if1\blind
{
	\title{\bf Model confidence sets and forecast combination: \hbox{An application to age-specific mortality}}
\maketitle
} \fi

\bigskip
\begin{abstract}

\underline{Background:} Model averaging combines forecasts obtained from a range of models, and it often produces more accurate forecasts than a forecast from a single model. 
\\

\underline{Objective:} The crucial part of forecast accuracy improvement in using the model averaging lies in the determination of optimal weights from a finite sample. If the weights are selected sub-optimally, this can affect the accuracy of the model-averaged forecasts. Instead of choosing the optimal weights, we consider trimming a set of models before equally averaging forecasts from the selected superior models. Motivated by \cite{HLN11}, we apply and evaluate the model confidence set procedure when combining mortality forecasts.
\\

\underline{Data \& Methods:} The proposed model averaging procedure is motivated by \cite{SS17} based on the concept of model confidence sets as proposed by \cite{HLN11} that incorporates the statistical significance of the forecasting performance. As the model confidence level increases, the set of superior models generally decreases. The proposed model averaging procedure is demonstrated via national and sub-national Japanese mortality for retirement ages between 60 and 100+. 
\\

\underline{Results:} Illustrated by national and sub-national Japanese mortality for ages between 60 and 100+, the proposed model-average procedure gives the smallest interval forecast errors, especially for males. 
\\

\underline{Conclusion:} We find that robust out-of-sample point and interval forecasts may be obtained from the trimming method. By robust, we mean robustness against model misspecification.

\vspace{.2in}
\end{abstract}

\newpage
\spacingset{1.5}

\section{Introduction}

Because of declining mortality rates in mainly developed countries, the improvement in human survival probability contributes significantly to an aging population. As a consequence, pension funds and insurance companies face longevity risk. The longevity risk is a potential risk attached to the increasing life expectancy of policyholders, which can eventually result in a higher payout ratio than expected \citep{CDR08}. The concerns about longevity risk have led to a surge in interest among pension funds and insurance companies in accurately modeling and forecasting age-specific mortality rates (or death counts or survival probabilities). Any improvement in the forecast accuracy of mortality rates will be beneficial for determining the allocation of current and future resources at the national and sub-national levels \citep[see, e.g.,][]{Koissia06, DDG07, HPG11}. 

Many different models for forecasting age-specific mortality rates have been proposed in the literature \citep[see][for reviews]{Booth06, BT08, CDE04, GK08, SBH11, TB14}. Of these, a significant milestone in demographic modeling and forecasting was the work by \cite{LC92}. They implemented a principal component method to model age-specific mortality and extracted a single time-varying index representing the trend in the level of mortality rates, from which the forecasts are obtained by a random walk with drift. While the Lee-Carter method is simple and robust in situations where age-specific log mortality rates have linear trends \citep{Booth06}, it has the limitation of attempting to capture the patterns of mortality rates using only one principal component and its associated scores. To rectify this deficiency, the Lee-Carter method has been extended and modified. For example, from a discrete data matrix perspective, \cite{BMS02}, \cite{RH03} and \cite{CBD06, CBD+09} proposed the use of more than one component in the Lee-Carter method to model age-specific mortality rates; \cite{RH06} proposed an age-period-cohort extension to the Lee-Carter model under a Poisson error structure, while \cite{Plat09b} extended this model by incorporating the dependence between ages. \cite{CBD06} used a logistic transformation to model the relationship between the death probability and age observed over time, while \cite{CBD+09} extended this model by incorporating the cohort effect. \cite{GK08} and \cite{WSB+15} considered a Bayesian paradigm for the Lee-Carter model estimation and forecasting. \cite{HH09} followed a Generalized Linear Model approach which leads to models that have a similar structure to the Lee-Carter model but with a generalized error structure. \cite{HB14} presented a general structural form of mortality models. From a continuous function perspective, \cite{HU07} proposed a functional data model that utilizes nonparametric smoothing and higher-order principal components, while \cite{SBH11} and \cite{Shang16} considered a multilevel functional data model to model mortality rates jointly for multiple populations.

There exist many papers on comparing the forecast accuracy among several mortality forecasting methods. However, the most accurate forecasting method has been determined based on an aggregate loss function. Instead of identifying the most accurate method, \cite{Shang12} considered a model averaging approach to combine forecasts from a range of methods, such as the Lee-Carter and functional time series methods. In this paper, we propose a new model averaging approach motivated by \cite{SS17}. The proposed model averaging method uses the model confidence set procedure to select a set of superior models and combines the forecasts by assigning equal weights to the set of superior models. In contrast with \cite{Shang12}, the problem is centered not on the selection of optimal weights, but on the selection of superior models. In Section~\ref{sec:MA_exist}, we compare the forecast accuracy between the existing and proposed model average methods.

With the aim of evaluating and comparing the forecast accuracy of different forecasting methods, forecast competition has a long history. The ``M" competition originated from \cite{MHM79} was the first attempt at a large empirical comparison of forecasting methods. In that ``M" competition, there were 1001 time series for which participants were invited to submit their forecasts. Later, the results were published in \cite{MAC+82}. The ``M" competition has progressed slowly over the years, with the most recent M4 competition taken place in 2018. The top-performing teams all combine forecasts from a range of statistical and machine learning methods via some model averaging to improve the point and interval forecast accuracies. 

Despite the popularity of model averaging in statistical and forecasting literature \citep[see, e.g.,][]{BG69, Dickinson75, Clemen89}, model averaging has not received increasing attention in the demographic literature with the noticeable exceptions of \cite{Shang12, Shang15} in the context of mortality forecasting, \citet[][Chapter 5]{Bijak11} in the context of migration forecasting, \cite{ABF+13} and \cite{SWB+14} for the overall population growth rate. \cite{Shang12} revisited many statistical methods and combined their forecasts based on two weighting schemes, one of which has been adapted for comparison in Section~\ref{sec:MA_exist}. Both weighting schemes determine the weights by using either in-sample forecast accuracy or in-sample goodness-of-fit. Because of the finite sample, the weight assigned to the worst model is often small but not zero. In turn, this may lead to inferior forecast accuracy than the one based on Oracle weights. This motivates us to consider an alternative model averaging idea. Instead of assigning weights to the forecasts from all models, we trim out the worse performing models based on a statistical significance test, such as the model confidence set procedure of \cite{HLN11} \citep[see also][]{SS17}.

While most attention has been paid to selecting model combination weights \citep[see, e.g.,][]{FH99, GKM+13}, \cite{ACT10} pointed out that there has been little research focusing on which models to include in the model pool. \cite{Graefe15} asserted that a simple average of the forecasts produced by individual models is a benchmark, and commonly outperforms more complicated weighting schemes that rely on the estimation of theoretically optimal weights. The simple average of the forecasts performs well when the model fits the data poorly; when the sample number per predictor is low and when the predictors are highly correlated \citep{Graefe15}.

This paper focuses on a statistical significance approach to select the models to be included in the forecast combination. Via the model confidence set procedure of \cite{HLN11}, we determine a set of statistically superior models, conditional on the model's in-sample performance for forecasting age-specific mortality rates. By equally averaging the forecasts from the superior models, we evaluate and compare point and interval forecast accuracies, as measured by the root mean square forecast error and mean interval score, respectively.

The outline of this paper is described as follows: From the Japanese national and sub-national age-specific mortality data in Section~\ref{sec:2}, we first visualize the heterogeneity in age-specific mortality rates among 47 prefectures. Then, we revisit some commonly used multivariate and functional time-series extrapolation methods for forecasting age-specific mortality rates in Section~\ref{sec:3}. Using the model confidence set procedure of \cite{HLN11} described in Section~\ref{sec:4}, we select a set of superior models based on their point or interval forecast accuracy and demonstrate the robust accuracy of the proposed model averaging method in Section~\ref{sec:5}. In Section~\ref{sec:MA_exist}, we present an adaption of an existing model-averaging method where optimal weights are estimated based on in-sample forecast accuracy and assigned to the forecasts from all models. In Section~\ref{sec:7}, we conclude and outline how the methodology presented here can be further extended.

\section{Japanese national and sub-national age-specific mortality data}\label{sec:2}

We study the Japanese age-specific mortality rates from 1975 to 2015, obtained from the \cite{JMD17}. We consider ages from 60 to 99 in single years of age, while the last age group contains all ages at and beyond 100 (abbreviated as 100+). We consider modeling mortality at older ages, as the mortality forecasts are an important input for calculating annuity prices for retirees and the corresponding reserves held by insurance companies and pension funds. Some of the models considered were designed for modeling mortality at older ages, such as the Cairns-Blake-Dowd models.

We split the Japanese mortality rates by sex and prefecture. We are also interested in the mortality data at the sub-national (i.e., prefecture) level. The mortality forecasts at the prefecture level are more useful than the mortality forecasts at the national level for local policy making and planning.

In the supplement, we have plotted the geographic locations (from North to South) of the 47 prefectures within eight regions of Japan. Also, we present the names of prefecture within each of the eight regions of Japan. \cite{SH17} and \cite{SH16} present plots of the ratio of mortality between each prefecture and Japan by age or year.

\section{Time-series extrapolation models}\label{sec:3}

We study some time-series extrapolation methods for modeling and forecasting age-specific mortality rates. The models that we have considered are subjective and far from extensive, but they suffice to serve as a test bed for demonstrating the performance of forecast combination. From actuarial science, we consider a family of Renshaw-Haberman (RH) models \citep[see, e.g.,][]{RH03, RH03b, RH06, RH08, HR08, HR09} and a family of Cairns-Blake-Dowd (CBD) models \citep{CBD06}. These two models perform well for mortality at higher ages, such as between 60 and 100+. From demography, we consider a family of Lee-Carter (LC) models \citep[see, e.g.,][]{LC92, BHT+06, Zhao12, ZLZ+13}. From statistics, we consider a family of functional time-series models \citep[see, e.g.,][]{HB08, HS09, HBY13}. For implementation, we use the StMoMo package of \cite{VVP18} for the RH and CBD models; we use the demography package of \cite{Hyndman17} for the LC models for the Gaussian error setting and functional time-series models. 

\subsection{Notations}

Let the random variable $D_{x,t}$ be the number of death counts in a population at age $x$ and year $t$. A rectangular data array $(d_{x,t}, e_{x,t})$ is available for data analysis where $d_{x,t}$ is the observed number of deaths and $e_{x,t}$ is the corresponding exposure to risk \citep{HH09}. The force of mortality and central mortality rates are given by $\mu_{x,t}$ and $m_{x,t} = d_{x,t}/e_{x,t}$, respectively. Cross-classification is by individual calendar year $t\in [t_1, t_n]$ (range $n$) and by age $x\in [x_1, x_k]$, either grouped into $k$ (ordered) categories, or by individual year (range $k$), in which case year-of-birth or cohort year $z=t-x \in [t_1-x_k, t_n-x_1]$ (range $n+k-1$) is defined \citep[see also][]{HH09}.

\subsection{Lee-Carter model under a Gaussian error setting}

With the central mortality rates $m_{x,t}$, the LC model structure is 
\begin{align}
m_{x,t} &= \exp^{\alpha_x + \beta_x^{(1)} \kappa_t^{(1)}+\varepsilon_{x,t}},\label{eq:LC} \\
\ln m_{x,t} &=\alpha_x + \beta_x^{(1)} \kappa_t^{(1)}+\varepsilon_{x,t} \notag
\end{align}
subject to the identification constraints
\begin{equation*}
\sum^{t_n}_{t=t_1}\kappa_t^{(1)} = 0, \quad \sum^{x_k}_{x=x_1}\beta_x^{(1)} = 1.
\end{equation*}
Note that $\alpha_x$ is the age pattern of the log mortality rates averaged across years; $\beta_x^{(1)}$ is the first principal component capturing relative change in the log mortality rate at each age $x$; $\kappa_t^{(1)}$ is the first set of principal component scores measuring general level of the log mortality rate at year $t$; bilinear terms $\beta_x^{(1)}\kappa_t^{(1)}$ incorporating the age-specific period trends \citep[][Section 6.2]{PDH+09}; and $\varepsilon_{x,t}$ is the model residual at age $x$ and year $t$.

In the demographic forecasting literature, the LC model adjusts $\kappa_t$ by refitting to the total number of deaths \citep[see][]{LC92}. In the \citeauthor{LM01}'s \citeyearpar{LM01} method, the adjustment of $\kappa_t$ involves fitting life expectancy at birth in the year $t$. In the \citeauthor{BHT+06}'s \citeyearpar{BHT+06} method, the adjustment of $\kappa_t$ involves fitting to the age distribution of deaths rather than to the total number of deaths. 

The adjusted principal component scores $\{\kappa_1, \dots, \kappa_n\}$ are then extrapolated by a random walk with drift method, from which forecasts are obtained by~\eqref{eq:LC} with the estimated mean function $\alpha_x$ and principal component $\beta_x$. That is,
\begin{equation*}
\ln\widehat{m}_{x,n+h|n} = \widehat{\alpha}_x + \widehat{\beta}_x\widehat{\kappa}_{n+h|n},
\end{equation*}
where $\widehat{\kappa}_{n+h|n}$ denotes forecasts of principal component scores obtained from a univariate time series forecasting method, such as the random walk with drift.

Two sources of uncertainty ought be considered: estimation errors in the parameters of the LC model and forecast errors in the forecast principal component scores. Because of orthogonality between the first principal component and the error term in~\eqref{eq:LC}, the overall forecast variance can be approximated by the sum of the two variances \citep[see also][]{LC92}. Conditioning on the past data $\bm{\mathcal{J}}=(\bm{m}_1,\dots,\bm{m}_n)$ and the first principal component $b_x$, we obtained the overall forecast variance of $\ln (m_{x, n+h})$,
\begin{equation}
\text{var}\left[\ln (m_{x, n+h})\big|\bm{\mathcal{J}}, b_x\right] \approx b_x^2 u_{n+h|n} + v_x,\label{eq:silly}
\end{equation}
where $b_x^2$ is the variance of the first principal component, calculated as the square of the $\beta_x$ in~\eqref{eq:LC}; $u_{n+h|n} = \text{var}(\kappa_{n+h}|\kappa_1,\dots,\kappa_n)$ can be obtained from the univariate time-series model; and the model residual variance $v_x$ is estimated by averaging the residual squares $\{\epsilon_{x,1}^2,\dots,\epsilon_{x,n}^2\}$ for each $x$ in~\eqref{eq:LC}.

\subsection{Renshaw-Haberman model under a Poisson error setting}
 
\cite{RH06} generalizes the LC model structure to include age-period-cohort modeling by formulating the mortality reduction factor as
\begin{equation}
\ln m_{x,t} = \alpha_x + \beta^{(1)}_x\kappa_t^{(1)} + \beta_x^{(0)}\gamma_{t-x}, \label{eq:RH}
\end{equation}
where $\alpha_x$ is an age function capturing the general shape of mortality by age; a time index $\kappa_t^{(1)}$ specifies the mortality trend and $\beta^{(1)}_x$ modulates its effect across ages; and $\gamma_{t-x}$ denotes a random cohort effect as a function of the birth-year $(t-x)$ \citep[see also][]{VVP18}. To estimate the parameters in~\eqref{eq:RH}, \cite{RH06} assume a Poisson distribution of deaths and use a log-link function targeting the force of mortality. 

To facilitate the model identifiability, a set of parameter constraints are imposed by setting
\begin{equation*}
\sum_{x} \beta_x^{(1)} = 1, \qquad \sum_{t}\kappa_t^{(1)} = 0, \qquad \sum_{x}\beta_x^{(0)} = 1, \qquad \sum_{c=t_1-x_k}^{t_n-x_1}\gamma_c = 0.
\end{equation*}

\subsubsection{Age-period-cohort (APC) model}

The APC model studied by \cite{CS87, CS87b} can be derived from the \cite{RH06} model. The APC model corresponds to $\beta_x^{(1)} = \beta_x^{(0)} = 1$ in~\eqref{eq:RH}, that is
\begin{equation*}
\ln m_{x,t} = \alpha_x + \kappa_t^{(1)} + \gamma_{t-x}.
\end{equation*}
To ensure the model identifiability, a set of parameters are constrained by setting
\begin{equation*}
\sum_t \kappa_t^{(1)} = 0, \qquad \sum_{c=t_1-x_k}^{t_n-x_1}\gamma_c = 0, \qquad \sum_{c=t_1 - x_k}^{t_n - x_1} c\gamma_c = 0.
\end{equation*}

\subsection{Cairns-Blake-Dowd (CBD) model}

While the LC model is a data-driven method, the CBD model attempts to find factors that may affect age-specific log mortality rates. The former approach is nonparametric, while the latter one is parametric. Note that in the original CBD model, the authors proposed the modeling of age-specific death probability $q_{x,t}$. Here, for the sake of comparison, we use the CBD model to model and forecast age-specific log mortality rates. Let the pre-specific age-modulating parameters be $\beta_x^{(1)} = 1$ and $\beta_x^{(2)} = x-\bar{x}$, the CBD model can be expressed as
\begin{equation}
\ln m_{x,t} = \kappa_t^{(1)} + (x-\bar{x})\kappa_t^{(2)},\label{eq:CBD}
\end{equation}
where $\bar{x}$ is the average age in the sample range. While $\kappa_t^{(1)}$ can be viewed as a time-varying interpret, $\kappa_t^{(2)}$ can be viewed as a time-varying slope. \cite{CBD06} produce mortality forecasts by projecting $\kappa_t^{(1)}$ and $\kappa_t^{(2)}$ jointly using a bivariate random walk with drift. 

\subsubsection{M7: Quadratic CBD model with cohort effects}

\cite{CBD+09} extend the original CBD model in~\eqref{eq:CBD} by adding a cohort effect and a quadratic age effect to form
\begin{equation*}
\ln m_{x,t} = \kappa_t^{(1)} + (x-\bar{x})\kappa_t^{(2)}+[(x-\bar{x})^2-\widehat{\sigma}_x^2]\kappa_t^{(3)} + \gamma_{t-x},
\end{equation*}
where $\widehat{\sigma}_x^2$ is the average value of $(x-\bar{x})^2$. To ensure the model identifiability, \cite{CBD+09} impose a set of constraints: 
\begin{equation*}
\sum_{c=t_1 - x_k}^{t_n - x_1}\gamma_c = 0, \qquad \sum_{c = t_1 - x_k}^{t_n - x_1}c\gamma_c = 0, \qquad \sum_{c=t_1 - x_k}^{t_n - x_1}c^2\gamma_c = 0.
\end{equation*}

In addition to M7 model, \cite{CBD+09} also consider two simpler predictors given by
\begin{align}
\ln m_{x,t} &= \kappa_t^{(1)} + (x-\bar{x})\kappa_t^{(2)}+\gamma_{t-x},\label{eq:M6}\\
\ln m_{x,t} &= \kappa_t^{(1)} + (x-\bar{x})\kappa_t^{(2)} + (x_c - x)\gamma_{t-x},\label{eq:M8}
\end{align}
where $x_c$ is a constant parameter to be estimated. Equations~\eqref{eq:M6} and~\eqref{eq:M8} are referred to as M6 and M8, respectively. 

\subsubsection{Plat model}

By combing the features of the LC and CBD models, \cite{Plat09} proposed the following model
\begin{equation}
\ln m_{x,t} = \alpha_x + \kappa_t^{(1)}+(\bar{x}-x)\kappa_t^{(2)}+(\bar{x}-x)^{+}\kappa_t^{(3)}+\gamma_{t-x}.
\end{equation}
where $(\bar{x}-x)^{+}=\max(\bar{x}-x, 0)$. To ensure the model identifiability, the following set of parameter constraints have been imposed
\begin{align*}
\sum_t \kappa_t^{(1)} = 0, \quad \sum_t \kappa_t^{(2)}=0,\quad \sum_t\kappa_t^{(3)}=0,\\
\sum_{c=t_1-x_k}^{t_n - x_1}\gamma_c = 0, \quad \sum_{c=t_1-x_k}^{t_n-x_1}c\gamma_c=0,\quad \sum_{c=t_1-x_k}^{t_n-x_1}c^2\gamma_c=0.
\end{align*}

In the families of the RH and CBD models, \cite{CBD06, CBD+11}, \cite{HR11} and \cite{VVP18} assume that the period indexes follow a multivariate random walk with drift. For the cohort index, they assume it follows a univariate autoregressive integrated moving average model. 

\subsection{The functional time-series models for one population}\label{sec:meme}

\subsubsection{Functional principal component analysis}

\cite{HU07} consider a functional time-series model for forecasting age-specific mortality rate, where age is treated as a continuum. The functional time-series model allows one to smooth the observed data points, in order to reduce or eliminate measurement error. To smooth data, \cite{HU07} suggest the use of a penalized regression spline with monotonic constraints applied to age-specific log mortality rates denoted by $\ln m_t(x)$ \citep[see][for detail]{HU07}. Here, we propose to smooth age-specific mortality rates from ages 0 to 100+, and then truncate the smoothed mortality rates from 60 to 100+. 

With smoothed age-specific log mortality curves ($\ln m_t(x)$), we obtain a mean function denoted by $\mu(x)$. With the de-centered smoothed data, we apply a functional principal component analysis to reduce dimensionality to some functional principal components (i.e, $\phi_k(x)$) and their associated scores $(\bm{\beta_k}=(\beta_{1,k},\dots,\beta_{n,k}))$. The functional principal component is constructed by sample variance of discretized functional data. Conditioning on the observed data and the estimated principal components, the point forecast of future mortality curves can be obtained by forecasting estimated principal component scores via a univariate time-series method. The prediction interval can be constructed similarly to the way of constructing prediction intervals for the Lee-Carter method in Section~\ref{eq:silly}. 

\subsubsection{Robust functional principal component analysis}

Because the presence of outliers can seriously affect the performance of modeling and forecasting, it is important to eliminate the effect of outliers where possible. As considered in \cite{HU07}, the robust functional time-series method calculates the integrated squared error for each year, that is
\begin{equation*}
\int_{\mathcal{I}} \Big[\ln m_t(x) - \mu(x) - \sum^K_{k=1}\beta_{t,k}\phi_k(x)\Big]^2 dx.
\end{equation*}
The integrated squared error provides a measure of estimation accuracy for the functional principal component approximation of the functional data. Outliers are those years that have a larger integrated squared error than a critical value calculated from a $\chi^2$ distribution \citep[see][for details]{HU07}. By assigning zero weight to outliers, we can again apply the functional time-series method to model and forecast age-specific mortality rates.

\subsection{The functional time-series models for multiple subpopulations}

When forecasting age-specific mortality for multiple subpopulations, it is advantageous to use a model that can capture correlation among subpopulations as the covariance of the multiple subpopulations often exhibits cross-correlation. By modeling the cross-correlation, it may improve forecast accuracy. Here, we consider the problem of jointly modeling and forecasting female and male mortality in order to produce coherent forecasts. Note that coherent forecasts can also be achieved by jointly modeling female or male mortality for all 47 prefectures.

\subsubsection{Product-ratio method of \cite{HBY13}}

In \cite{HBY13}, they define the square roots of the product and ratio functions of the smoothed mortality rates for female and male data:
\begin{align*}
p_t(x) &= \sqrt{m_{t}^{\text{M}}(x)m_{t}^{\text{F}}(x)}, \\ 
r_t(x) &= \sqrt{m_{t}^{\text{M}}(x)/m_{t}^{\text{F}}(x)}.
\end{align*}
Instead of modeling female and male mortality data, we model the product and ratio functions. The advantage of this approach is that the product and ratio functions tend to behave roughly independently of each other, provided that the multiple subpopulations have approximately equal variances. On the logarithmic scale, these are sums and differences that are nearly uncorrelated. The functional time-series method in Section~\ref{sec:meme} can be applied to forecast the product and ratio functions \citep[see][fore details]{HBY13}.

\subsubsection{Multivariate functional time-series method}

We consider data where each observation consists of $w\geq 2$ functions, $[\ln m^{(1)}(x),\dots, \ln m^{(w)}(x)]^{\top}\in R^{w}$. These multivariate functions are defined over the same domain $\mathcal{I}$ \citep[e.g.,][]{JP14, CCY14, SY17}.

We follow \cite{SY17} and consider the stacking of multiple subpopulations into a long vector of functions, i.e., we stack the discretized data points of each sub-population together for the same year. Then, we perform a multivariate functional principal component analysis to reduce dimensionality and summarize the main mode of information. With the extracted principal components and their scores, a functional time-series method in Section~\ref{sec:meme} can be applied again.

\subsubsection{Multilevel functional time-series method}

The multilevel functional data model has a strong resemblance to a two-way functional analysis of variance model studied by \cite{MVB+03, CF10} and \cite[][Section 5.4]{Zhang14}. It is a special case of the general `functional mixed model' proposed in \cite{MC06}. In the case of two subpopulations, the basic idea is to decompose age-specific log mortality curves among different subpopulations into a sex-specific average $\mu^j(x)$, a common trend across subpopulations $R_t(x)$, a sex-specific residual trend $U_t^j(x)$, and measurement error $e_t^j(x)$ with finite variance $(\sigma^2)^j$ \citep[see, e.g.,][]{HH13, Shang16}. The common and sex-specific residual trends are modeled by projecting them onto the eigenvectors of covariance operators of the aggregate and population-specific centered stochastic processes, respectively. 

\section{Model confidence set}\label{sec:4}

The model confidence set procedure proposed by \cite{HLN11} consists of a sequence of tests permitting the construction of a set of ``superior" models, where the null hypothesis of equal predictive ability (EPA) is not rejected at a specified confidence level. The EPA test statistic can be evaluated for any arbitrary loss function, such as the square or absolute loss function. 

Let $M$ be some subset of original models denoted by $M^0$ and let $m$ be the number of models in $M$, and let $d_{\rho \xi, \ell}$ denote the loss differential between two models $\rho$ and $\xi$, that is
\begin{equation*}
d_{\rho \xi, \ell} = l_{\rho, \ell} - l_{\xi, \ell}, \qquad \rho, \xi=1,\dots,m, \quad \ell=1,\dots,N,
\end{equation*}
and calculate 
\begin{align*}
d_{\rho\cdot, \ell} = \frac{1}{m}\sum_{\xi\in M}d_{\rho \xi,\ell}, \qquad \rho = 1,\dots,m
\end{align*}
as the loss of model $\rho$ relative to any other model $\xi$ at time point $\ell$. Let $c_{\rho \xi} = \text{E}(d_{\rho \xi})$ and $c_{\rho.} = \text{E}(d_{\rho.})$ be finite and not time dependent. The EPA hypothesis for a set of $M$ candidate models can be formulated in two ways:
\begin{align}
\text{H}_{0,\text{M}}: c_{\rho \xi}&=0, \qquad \text{for all}\quad \rho, \xi = 1,2,\dots,m\notag\\
\text{H}_{\text{A,M}}: c_{\rho \xi}&\neq 0, \qquad \text{for some}\quad \rho, \xi = 1,2,\dots,m.\label{eq:hypo_1}
\end{align}
or
\begin{align}
\text{H}_{0,\text{M}}: c_{\rho.}&=0, \qquad \text{for all}\quad \rho = 1,2,\dots,m\notag\\
\text{H}_{\text{A,M}}: c_{\rho.}&\neq 0, \qquad \text{for some}\quad \rho = 1,2,\dots,m.\label{eq:hypo_2}
\end{align}
Based on $c_{\rho \xi}$ or $c_{\rho.}$, we construct two hypothesis tests as follows:
\begin{align}
t_{\rho \xi} &= \frac{\overline{d}_{\rho \xi}}{\sqrt{\widehat{\text{Var}}(\overline{d}_{\rho \xi})}} \label{eq:t_ij_t_i_1},\\
t_{\rho.} &= \frac{\overline{d}_{\rho.}}{\sqrt{\widehat{\text{Var}}(\overline{d}_{\rho.})}}, \label{eq:t_ij_t_i}
\end{align}
where $\overline{d}_{\rho.} = \frac{1}{m}\sum_{\xi\in M}\overline{d}_{\rho \xi}$ is the sample loss of the $\rho^{\text{th}}$ model compared to the averaged loss across models, and $\overline{d}_{\rho \xi} =\frac{1}{m}\sum^m_{\ell=1}d_{\rho\xi,\ell}$ measures the relative sample loss between the $\rho^{\text{th}}$ and $\xi^{\text{th}}$ models. Note that $\widehat{\text{Var}}\left(\overline{d}_{\rho.}\right)$ and $\widehat{\text{Var}}\left(\overline{d}_{\rho\xi}\right)$ are the bootstrapped estimates of $\text{Var}\left(\overline{d}_{\rho.}\right)$ and $\text{Var}\left(\overline{d}_{\rho\xi}\right)$, respectively. \cite{BC14} perform a block bootstrap procedure with 5,000 bootstrap samples by default, where the block length is given by the maximum number of significant parameters obtained by fitting an autoregressive process on all the $d_{\rho\xi}$ terms. For both hypotheses in~\eqref{eq:hypo_1} and~\eqref{eq:hypo_2}, there exist two test statistics:
\begin{align*}
T_{\text{R}, \text{M}} = \max_{\rho,\xi\in M}\left|t_{\rho\xi}\right|,\qquad T_{\max, \text{M}} = \max_{\rho\in M} t_{\rho.},
\end{align*} 
where $t_{\rho\xi}$ and $t_{\rho.}$ are defined in~\eqref{eq:t_ij_t_i_1} and~\eqref{eq:t_ij_t_i}, respectively. While $T_{\text{R}, \text{M}}$ uses the loss differential between models $\rho$ and $\xi$, $T_{\max, \text{M}}$ uses the aggregated loss differential between models $\rho$ and $\xi$ over $\xi$. Oftentimes but not always, the models selected on the basis of $T_{\text{R}, \text{M}}$ form a subset of the models selected on the basis of $T_{\max, \text{M}}$.
 
The Model Confidence Set (MCS) procedure is a sequential testing procedure, which eliminates the worst model at each step until the hypothesis of equal predictive ability is accepted for all the models belonging to a set of superior models. The selection of the worst model is determined by an elimination rule that is consistent with the test statistic, 
\begin{align*}
e_{\text{R},\text{M}} = \argmax_{\rho\in M}\left\{\sup_{\xi \in M}\frac{\overline{d}_{\rho\xi}}{\sqrt{\widehat{\text{Var}}\left(\overline{d}_{\rho\xi}\right)}}\right\}, \qquad
e_{\max, \text{M}} = \argmax_{\rho\in M}\frac{\overline{d}_{\rho.}}{\widehat{\text{Var}}\left(\overline{d}_{\rho.}\right)}.
\end{align*} 

\section{Forecast results}\label{sec:5}

\subsection{Point forecast evaluation}

An expanding window analysis of a time series model is commonly used to assess model and parameter stabilities over time. It assesses the constancy of a model's parameter by computing parameter estimates and their corresponding forecasts over an expanding window of a fixed size through the sample size \citep[see][Chapter 9 for details]{ZW06}. Using the first 21 observations from 1975 to 1995 in the Japanese age-specific mortality rates, we produce one-step-ahead point forecasts. Through an expanding window approach, we re-estimate the parameters in the time series forecasting models using the first 22 observations from 1975 to 1996. Forecasts from the estimated models are then produced for one-step-ahead. We iterate this process by increasing the sample size by one year until reaching the end of the training data period in 2005. This process produces ten one-step-ahead forecasts in the validation data period from 1996 to 2005. We compare these forecasts with the holdout samples to determine the point and interval forecast accuracies. By using the MCS procedure, we identify a superior set of models for averaging. Through the expanding window approach, we evaluate the out-of-sample point and interval forecast accuracies for the testing data from 2006 to 2015.

To evaluate the point forecast accuracy, we use the root mean squared forecast error (RMSFE). The RMSFE measures how close the forecasts are in comparison to the actual values of the variable being forecast. For each series, they can be written as
\begin{equation*}
\text{RMSFE}_{\xi} = \sqrt{\frac{1}{41}\sum^{41}_{j=1}\left[\mathcal{Y}_{n+\xi}(x_j) - \widehat{\mathcal{Y}}_{n+\xi}(x_j)\right]^2},
\end{equation*}
where $\mathcal{Y}_{n+\xi}(x_j)$ represents the actual holdout sample for the $j$\textsuperscript{th} age and $\xi$\textsuperscript{th} curve of the forecasting period, while $\widehat{\mathcal{Y}}_{n+\xi}(x_j)$ represents the point forecasts for the holdout sample.

\subsection{Interval forecast evaluation}

In addition to point forecasts, we also evaluate the pointwise interval forecast accuracy using the interval score of \cite{GR07} \citep[see also][]{GK14}. For each year in the forecasting period, the one-step-ahead prediction intervals were calculated at the $100(1-\alpha)\%$ nominal coverage probability. We consider the common case of symmetric $100(1-\alpha)\%$ prediction interval, with lower and upper bounds that are predictive quantiles at $\alpha/2$ and $1-\alpha/2$, denotes by $\widehat{\mathcal{Y}}_{n+\xi}^l(x_i)$ and $\widehat{\mathcal{Y}}_{n+\xi}^u(x_i)$. As defined by \cite{GR07}, a scoring rule for the \textit{pointwise} interval forecast at time point $x_i$ is
\begin{align*}
S_{\alpha}\left[\widehat{\mathcal{Y}}_{n+\xi}^l(x_i), \widehat{\mathcal{Y}}_{n+\xi}^u(x_i); \mathcal{Y}_{n+\xi}(x_i)\right] = \left[\widehat{\mathcal{Y}}_{n+\xi}^u(x_i) - \widehat{\mathcal{Y}}_{n+\xi}^l(x_i)\right] + \frac{2}{\alpha}\left[\widehat{\mathcal{Y}}_{n+\xi}^l(x_i)-\mathcal{Y}_{n+\xi}(x_i)\right] \\
\mathds{1}\left\{\widehat{\mathcal{Y}}_{n+\xi}(x_i) < \mathcal{Y}_{n+\xi}(x_i)\right\} + \frac{2}{\alpha} \left[\mathcal{Y}_{n+\xi}(x_i) - \widehat{\mathcal{Y}}_{n+\xi}^u(x_i)\right]\mathds{1}\left\{\mathcal{Y}_{n+\xi}(x_i) > \widehat{\mathcal{Y}}_{n+\xi}(x_i)\right\},
\end{align*}
where $\alpha$ denotes the level of significance, customarily $\alpha=0.2$; and $\mathds{1}\{\cdot\}$ denotes a binary indicator. The optimal interval score is achieved when $\mathcal{Y}_{n+\xi}(x_i)$ lies between $\widehat{\mathcal{Y}}_{n+\xi}^l(x_i)$ and $\widehat{\mathcal{Y}}_{n+\xi}^u(x_i)$, with the distance between the upper bound and lower bound being minimal. 

We define the mean interval score for different points in a curve and different lengths in the forecasting period as
\begin{equation}
\overline{S}_{\alpha, \xi} = \frac{1}{41}\sum^{41}_{j=1}S_{\alpha}\left[\widehat{\mathcal{Y}}_{n+\xi}^l(x_j), \widehat{\mathcal{Y}}_{n+\xi}^u(x_j); \mathcal{Y}_{n+\xi}(x_j)\right],
\end{equation}
where $S_{\alpha, \xi}\left[\widehat{\mathcal{Y}}_{n+\xi}^l(x_j), \widehat{\mathcal{Y}}_{n+\xi}^u(x_j); \mathcal{Y}_{n+\xi}(x_j)\right]$ denotes the interval score at the $\xi$\textsuperscript{th} curve of the forecasting period.

\subsection{Determining a superior set of models}

Based on the RMSFE error measure in the training data, we examine statistical significance in point forecast accuracy among the 17 time-series extrapolation methods. The 17 models considered are listed in Table~\ref{tab:method}.
\begin{table}[!htbp]
\centering
\tabcolsep 0.14in
\caption{A list of the 19 models considered.}\label{tab:method}
\begin{tabular}{@{}lllll@{}}
\toprule
Family of models & & Label & & Model \\
\midrule
Renshaw-Haberman & & 1 & & Lee-Carter model with Poisson error structure \\
	& & 2 & & Renshaw-Haberman model \\
	& & 3 & & Age-period-cohort model \\ 
\\	
Cairns-Blake-Dowe 	& & 4 & & Cairns-Blake-Dowe model \\
	& & 5 & & M6 model \\
	& & 6 & & M7 model \\
	& & 7 & & M8 model \\
	& & 8 & & Plat model \\
	\\
Lee-Carter & & 9 & & Lee-Carter model with Gaussian error structure \\
	& & 10 & & Booth-Maindonald-Smith model \\
	& & 11 & & Lee-Carter model with adjustment of life expectancy \\
	& & 12 & & Lee-Carter model with no adjustment to the score \\
	\\
Functional time series & & 13 & & Functional data model \\
	& & 14 & & Robust functional data model \\
	& & 15 & & Product-ratio model \\
	& & 16 & & Multivariate functional data model \\
	& & 17 & & Multilevel functional data model \\	
	\\
Model averaging & & 18 & & MCS ($T_{\max, \text{M}}$) to select equal-weighted models  \\
			& & 19 & & MCS ($T_{\text{R}, \text{M}}$) to select equal-weighted models  \\	
\bottomrule
\end{tabular}
\end{table}

With the 90\% confidence level of the MCS tests, we identify the set of superior models regarding point forecast accuracy. In Tables~\ref{tab:24} and~\ref{tab:24b}, we determine the set of superior models among the 17 models considered using the $T_{\max, \text{M}}$ and $T_{\text{R}, \text{M}}$ tests. 

\begin{table}[!htbp]
\begin{small} 
\begin{center}
\tabcolsep 0.4cm
\caption{MCS procedure using the $T_{\max, \text{M}}$ test applied to the RMSFE in the validation set from 1996 to 2005 for forecasting the Japanese female and male national and sub-national mortality for ages between 60 and 100+. From the 17 models, below is the selected superior set of the model(s).} \label{tab:24} 
\begin{tabular}{@{}lll@{}}
\toprule
Population & \multicolumn{2}{c}{Superior models} \\
 & Female  & Male \\\midrule
Japan & 1 & 1 \\
Hokkaido & 1, 8, 9, 10, 11, 12, 13, 14, 15, 16 &  16 \\
Aomori & 7 & 16 \\
Iwate & 3, 5, 7, 8, 13, 14, 15, 17 & 13 \\
Miyagi & 2, 3, 5, 7, 8, 9, 10, 11, 12, 13, 14, 15, 16, 17 & 1, 3, 5, 8, 13, 14, 15, 16, 17 \\
Akita & 7 & 14 \\
Yamagata & 13, 14, 16 &  5, 7 \\
Fukushima & 3 & 13 \\
Ibaraki & 1, 2, 3, 5, 7, 9, 10, 11, 12, 13, 14, 15, 17 & 16 \\
Tochigi & 2, 7, 14, 15 & 8, 17 \\
Gunma & 8 & 2, 3, 5, 7, 8, 9, 10, 11, 12, 13, 14, 15, 16, 17 \\
Saitama & 8 & 14 \\
Chiba & 3, 8, 9, 10, 11, 15 & 16 \\
Tokyo & 8 & 13, 15, 16 \\
Kanagawa & 8 &  14 \\
Niigata & 3, 5, 7, 8, 14, 15 & 3, 5, 7, 8, 13, 17 \\
Toyama & 3, 5, 7, 8 & 13, 17 \\
Ishikawa & 2, 3, 4, 5, 7, 8, 13, 14, 15, 17 & 1, 2, 3, 4, 5, 6, 7, 8, 9, 10, 11, 12, 13, 14, 15, 16, 17 \\
Fukui & 3 & 3, 4, 5, 7, 8, 13, 14, 15, 16, 17 \\
Yamanashi & 7 &  5, 13, 14, 15, 16, 17 \\
Nagano & 8 & 14 \\
Gifu & 13 & 3, 5, 7, 13, 14, 15, 16, 17 \\
Shizuoka & 1, 8, 10 & 13, 14, 15, 16 \\
Aichi & 1, 3, 8, 9, 10, 11, 12, 14 & 13, 14, 15, 17 \\
Mie & 9 & 8 \\
Shiga & 8 & 13 \\
Kyoto & 8, 13, 14, 15, 17 & 14 \\
Osaka & 8 & 17 \\
Hyogo & 8 & 13, 14, 15, 16, 17 \\
Nara & 3, 5, 7, 8, 13, 14 & 13, 14, 15, 16, 17 \\
Wakayama & 8 & 17 \\
Tottori & 7 & 3, 5, 8, 13, 14, 15, 16, 17 \\
Shimane & 13, 15 & 1, 3, 5, 7, 8, 13, 14, 15, 16, 17 \\
Okayama & 10 & 1, 3, 8, 13, 14, 15, 16 \\
Hiroshima & 10 & 15 \\
Yamaguchi & 8, 15 & 14 \\
Tokushima & 7 & 7 \\
Kagawa & 14 & 3, 5, 7, 8, 13, 14, 15, 16, 17 \\
Ehime & 7 & 3, 8, 13, 14, 15, 16, 17 \\
Kochi & 14 & 17 \\
Fukuoka & 1 & 13 \\
Saga & 3, 5, 7, 8, 9, 10, 11, 12, 15 & 5, 7, 14 \\
Nagasaki & 1, 8, 10, 11, 13, 14, 15, 16, 17 & 17 \\
Kumamoto & 15 & 14 \\
Oita & 13, 14 & 5, 7, 13, 14, 15, 16 \\
Miyazaki & 2, 3, 5, 8, 13, 14, 15, 17 & 16 \\ 
Kagoshima & 1, 3, 8, 9, 10, 11, 12, 13, 14, 15 & 3, 13, 14 \\
Okinawa & 9, 10, 11, 12, 13, 14, 15, 17 & 13, 14, 15, 16, 17 \\
\bottomrule
\end{tabular}
\end{center} 
\end{small}
\end{table}

\begin{table}[!htbp]
\begin{small}
\begin{center}
\tabcolsep 2cm
\caption{MCS procedure using the $T_{\text{R}, \text{M}}$ test applied to the RMSFE in the validation set from 1996 to 2005 for forecasting the Japanese female and male national and sub-national mortality for ages between 60 and 100+.} \label{tab:24b} 
\begin{tabular}{@{}lll@{}}
\toprule
Population & \multicolumn{2}{c}{Superior models} \\
 & Female  & Male \\\midrule
Japan & 1 & 1 \\
Hokkaido & 1 & 16 \\
Aomori & 	7 & 16 \\
Iwate & 8 & 13 \\
Miyagi & 2, 3, 7, 8, 9, 10, 12, 14, 15 & 3, 13, 14, 15, 16, 17 \\
Akita & 7 & 14 \\
Yamagata & 13, 14, 16 &  5, 7, 13, 15, 16 \\
Fukushima & 3 & 13 \\
Ibaraki & 3, 7, 14, 15 & 16 \\
Tochigi & 15 & 8, 17 \\
Gunma & 8 & 2, 3, 5, 8, 13, 14 \\
Saitama & 8 & 14 \\
Chiba & 3, 8, 9 & 16 \\
Tokyo & 8 & 16 \\
Kanagawa & 8 & 14 \\  
Niigata & 3, 7, 8, 14 & 3, 5, 8, 13, 17 \\ 
Toyama & 3, 5, 7, 8 & 13, 17 \\
Ishikawa & 5, 7, 13 & 3, 5, 7, 13, 17 \\
Fukui & 3 & 3, 7, 8, 13, 14, 16, 17 \\
Yamanashi & 7 &  15 \\
Nagano & 8 & 14 \\
Gifu & 13 & 	3, 5, 7, 13, 14, 15, 17 \\
Shizuoka & 1, 8, 10 & 8, 13, 14, 15, 16 \\
Aichi & 8 & 15 \\
Mie & 3, 5, 7, 8, 9, 10, 11 & 8 \\
Shiga & 8 & 13 \\
Kyoto &  14 & 14 \\
Osaka & 8 & 17 \\
Hyogo & 8 & 15 \\
Nara & 3, 5, 7 & 14 \\	
Wakayama & 	8 & 17 \\
Tottori & 7 & 3, 5, 8, 16, 17 \\
Shimane & 13, 15 & 8, 15, 16, 17 \\
Okayama & 10 & 1, 3, 8, 13 \\
Hiroshima & 10 & 15 \\
Yamaguchi & 8, 15 & 14 \\
Tokushima & 3, 5, 7, 8, 15 & 5, 7, 13, 14, 15, 16 \\ 
Kagawa & 14 & 3, 5, 7, 13, 15, 17 \\
Ehime & 7 & 1, 3, 8, 13, 14, 15, 16, 17 \\
Kochi & 14 & 17 \\
Fukuoka & 1, 9, 10 & 13 \\
Saga & 3, 7, 8, 11, 15 & 7 \\
Nagasaki & 13 & 17 \\
Kumamoto & 10, 11, 12, 13, 14, 15, 17 & 14 \\
Oita & 13, 14, 17 & 5, 7, 13 \\
Miyazaki & 3 & 16 \\
Kagoshima & 1, 3, 8, 10, 11, 12, 14, 15 & 3, 13 \\
Okinawa & 12, 14 & 15, 16, 17 \\
\bottomrule
\end{tabular}
\end{center} 
\end{small}
\end{table}

Based on the mean interval scores in the validation period, we examine statistical significance in interval forecast accuracy among the time-series extrapolation methods. With the 90\% confidence level of the MCS tests, we identify the set of superior models regarding interval forecast accuracy. In Tables~\ref{tab:240} and~\ref{tab:240b}, we determine the set of superior models using both the $T_{\max, \text{M}}$ and $T_{\text{R}, \text{M}}$ tests among the 17 models considered.

\begin{table}[!htbp]
\begin{small}
\begin{center}
\tabcolsep 1.9cm
\caption{MCS procedure using the $T_{\max, \text{M}}$ test applied to the mean interval score in the validation set from 1996 to 2005 for forecasting the Japanese female and male national and sub-national mortality rates for ages between 60 and 100+.} \label{tab:240} 
\begin{tabular}{@{}lll@{}}
\toprule
Population & \multicolumn{2}{c}{Superior models} \\
 & Female & Male \\\midrule
Japan & 8 & 8 \\
Hokkaido & 8, 13, 14, 15, 16 & 15 \\
Aomori & 8 & 14 \\
Iwate & 13 & 15 \\
Miyagi & 15 & 15 \\
Akita & 17 & 15 \\
Yamagata & 13 & 15 \\
Fukushima & 15 & 15 \\
Ibaraki & 8, 13, 14, 15, 17 & 15 \\
Tochigi & 15 & 15 \\
Gunma & 8 & 15 \\
Saitama & 8 & 15 \\
Chiba & 8 & 17 \\
Tokyo & 8 & 15 \\
Kanagawa & 8 & 15 \\
Niigata & 8 & 15, 17 \\
Toyama & 8 & 15 \\
Ishikawa & 7, 8, 13, 14, 15, 17 & 13, 14, 15, 17 \\
Fukui & 8 & 13 \\
Yamanashi & 8 & 15 \\
Nagano & 8 & 15 \\
Gifu & 15 & 15 \\
Shizuoka & 8 & 15 \\
Aichi & 8 & 15 \\
Mie & 8, 13, 14, 15, 16 & 15 \\
Shiga & 8 & 15 \\
Kyoto & 14, 15 & 15 \\
Osaka & 8 & 15 \\
Hyogo & 1, 3, 5, 7, 8, 9, 10, 11, 12, 14, 15, 16, 17 & 17 \\
Nara & 8, 13, 14, 15, 16 & 15 \\
Wakayama & 8, 15 & 15 \\
Tottori & 8 & 15 \\
Shimane & 15 & 15 \\
Okayama & 15 & 15 \\
Hiroshima & 8 & 15 \\
Yamaguchi & 8 & 15 \\
Tokushima & 15, 16 & 13, 14, 15, 16 \\
Kagawa & 8, 13, 14, 15 & 15 \\
Ehime & 8 & 15 \\
Kochi & 17 & 15 \\
Fukuoka & 8 & 15 \\
Saga & 15 & 15 \\
Nagasaki & 8 & 15 \\
Kumamoto & 13, 14, 15, 17 & 15 \\
Oita & 15 & 15 \\
Miyazaki & 17 & 15 \\
Kagoshima & 8, 13, 14, 15, 16 & 15 \\
Okinawa & 15 & 15 \\
\bottomrule
\end{tabular}
\end{center} 
\end{small}
\end{table}

\begin{table}[!htbp]
\begin{small}
\begin{center}
\tabcolsep 2.9cm
\caption{MCS procedure using the $T_{\text{R}, \text{M}}$ test applied to the mean interval score in the validation set from 1996 to 2005 for forecasting the Japanese female and male national and sub-national mortality rates for ages between 60 and 100+.} \label{tab:240b} 
\begin{tabular}{@{}lll@{}}
\toprule
Population & \multicolumn{2}{c}{Superior models} \\
 & Female & Male \\\midrule
Japan & 8 & 8 \\
Hokkaido & 8, 13, 15, 16, 17 & 15 \\
Aomori & 8 & 14 \\
Iwate & 13 & 15 \\
Miyagi & 15 & 15 \\
Akita & 17 & 15 \\
Yamagata & 13 & 15 \\
Fukushima & 15 & 15 \\
Ibaraki & 8, 13, 14, 17 & 15 \\
Tochigi & 15 & 15 \\
Gunma & 8 & 15\\
Saitama & 8 & 15 \\
Chiba & 8 & 17\\
Tokyo & 8 & 15 \\
Kanagawa & 8 & 15 \\
Niigata & 8 & 13, 15, 17 \\
Toyama & 8 & 15\\
Ishikawa & 13 & 15\\
Fukui & 8 & 13\\
Yamanashi & 8 & 15\\
Nagano & 8 & 15\\
Gifu & 8 & 15\\
Shizuoka & 8 & 15\\
Aichi & 8 & 15\\
Mie & 8, 13, 15 & 15\\
Shiga & 8 & 15\\
Kyoto & 14, 15 & 15\\
Osaka & 8 & 15\\
Hyogo & 8 & 17\\
Nara & 8, 15, 16 & 15\\
Wakayama & 8, 15 & 15\\
Tottori & 8 & 15\\
Shimane & 15 & 15\\
Okayama & 15 & 15\\
Hiroshima & 8 & 15\\
Yamaguchi & 8 & 13, 14, 15\\
Tokushima & 15, 16 & 13, 15, 16\\
Kagawa & 8, 14, 15 & 15, 16, 17\\
Ehime & 8 & 15\\
Kochi & 17 & 15\\
Fukuoka & 8 & 15\\
Saga & 15 & 15\\
Nagasaki & 8 & 15\\
Kumamoto & 13, 15, 17 & 15\\
Oita & 15 & 15\\
Miyazaki & 17 & 15\\
Kagoshima & 15 & 15\\
Okinawa & 15 & 15\\
\bottomrule
\end{tabular}
\end{center} 
\end{small}
\end{table}

\subsection{Point and interval forecast comparison}

Based on the selected set of superior models, we produce model-averaged point and interval forecasts using equal weights. In Table~\ref{tab:4}, we compute the point and interval forecast accuracies for female and male data. 

\begin{table}[!htbp]
\caption{Point and interval forecast accuracies among the 17 models and two model-averaged methods in the Japanese national data and the average of 47 sub-national populations for ages between 60 and 100+. Forecast errors have been multiplied by 100. The smallest overall errors are shown in bold.}\label{tab:4} 
\centering
\tabcolsep 0.1in
\begin{tabular}{@{}llrrrrr@{}}
\toprule
& & \multicolumn{2}{c}{RMSFE} & & \multicolumn{2}{c}{Mean interval score} \\ 
Series & Method & National data & Sub-national data &  & National data & Sub-national data \\
\midrule
Female & 1 & 0.54 & 1.11 & & 1.81 & 3.88 \\
	     & 2 & 4.20 & 6.24 & & 7.12 & 61.70 \\
	    & 3 & 1.34 & 1.61 & &  4.91 & 5.41\\
	    & 4 & 1.98 & 2.17 & & 7.24  & 7.09\\
	    & 5 & 1.20 & 1.51 & & 3.71  & 4.50    \\
	    & 6 & 2.05 & 2.27 & & 5.95 & 5.84	\\
	    & 7 & 0.87 & 1.27 & & 1.93 & 2.89 	\\
	    & 8 & \textBF{0.33} & \textBF{1.02} & & \textBF{0.86} & \textBF{2.40} 	\\
	    & 9 & 0.67 & 1.26 & & 2.10 & 4.26	\\
	    & 10 & 0.69 & 1.27 & & 2.24 & 4.34	\\
	    & 11 & 0.69 & 1.27 & & 2.20 & 4.33	\\
	    & 12 & 0.69 & 1.28 & & 2.22 & 4.30	\\
	    & 13 & 0.71 & 1.21 & & 1.31 & 2.71	\\
	    & 14 & 0.72 & 1.21 & & 1.35 & 2.70	\\
	    & 15 & 0.61 & 1.21 & & 1.32 & 2.94	\\
	    & 16 & 0.83 & 1.25 & & 1.43 & 3.40	\\
	    & 17 & 0.81 & 1.23 & & 1.80 & 2.73	\\
	    \cmidrule{2-7}
& 18 & 0.54 & 1.24 & & 0.87 & 2.57	\\
& 19 & 0.54 & 1.22 & & 0.87 & 2.57	\\
\midrule
Male & 1 & 0.71 & 2.55 & & 2.79 & 9.56	\\
	& 2 & 2.27 & 4.02 & & 7.97 & 14.57	\\
	& 3 & 1.93 & 3.13 & & 8.17 & 11.40	\\
	& 4 & 2.92 & 3.76 & & 12.18 & 13.01	\\
	& 5 & 1.78 & 2.96 & & 6.42	 & 9.45 \\
	& 6 & 2.75 & 3.69 & & 9.72	 & 10.47 \\
	& 7 & 1.60 & 2.84 & & 3.81	 & 7.91 \\
	& 8 & \textBF{0.65} & 2.47 & & 1.37	 & 5.77 \\
	& 9 & 0.99 & 3.78 & & 3.68	 & 12.80 \\
	& 10 & 1.00 & 3.76 & & 3.80 & 12.78	\\
	& 11 & 1.02 & 3.83 & & 3.91 & 13.05	\\
	& 12 & 1.00 & 3.47 & & 3.63 & 10.88	\\
	& 13 & 0.72 & 2.51 & & 1.50 & 5.86	\\
	& 14 & 0.78 & 2.51 & & 1.49 & 5.87	\\
	& 15 & 0.72 & 2.47 & & 1.98 & \textBF{5.02}	\\
	& 16 & 0.93 & 2.47 & & 1.79 & 7.23	\\
	& 17 & 0.76 & \textBF{2.44} & & 1.79 & 5.49	\\
	\cmidrule{2-7}
& 18 & 0.71 & 2.51 & & \textBF{1.36} &5.06	\\
& 19 	& 0.71 & 2.50 & & \textBF{1.36} & 5.07	\\
\bottomrule
\end{tabular}
\end{table}

As measured by the mean RMSFE over ten years in the forecasting period, the Plat model gives the most accurate point forecasts for the whole of Japan. For the average of 47 prefectures in Japan, the Plat model gives the most accurate point forecasts for females, while the multilevel functional time-series method performs the best for males. Although the model-averaging methods are not the best model, in this case, they rank among the top-performing methods. Between the two statistical significance tests, there is a marginal difference in the point forecast accuracy.

As measured by the mean interval scores over ten different horizons, the Plat model produces the most accurate interval forecasts for the whole of Japan and its prefectures for females. For males, the model-averaging methods perform the best for the whole of Japan and rank as the second best after the product-ratio method for the average of 47 prefectures. Again, there is a marginal difference between the two statistical significance tests regarding interval forecast accuracy.

\section{A competing model averaging method}\label{sec:MA_exist}

An existing model averaging method combines forecasts from the top two methods \citep[see][]{Shang12}. Given the top two methods are arbitrary from sample to sample, we have decided to combine forecasts from all methods and assign weights differently. Among all methods, we determine point or interval forecast accuracy as measured by the corresponding forecast errors in the validation set, and assign the weights to be the inverse of their forecast errors. We then standardize all weights so that the weights sum to 1. Conceptually, the method that performs better in the validation set receives a higher weight in the combined forecasts. The point and interval forecast accuracies of this model-averaging method are presented in Table~\ref{tab:MA_exist}.

\begin{table}[ht]
\centering
\tabcolsep 0.12in
\caption{Point and interval forecast accuracies for an existing model-averaged method averaged across the Japanese female and male national and sub-national mortality rates for ages between 60 and 100+. Forecast errors have been multiplied by 100. The results show its inferior point and interval forecast accuracies compared to the proposed two model-averaging methods. This further confirms that one should not average all models, but a subset of all `good' models.}\label{tab:MA_exist}
\begin{tabular}{@{}lrrrrrrrrrrr@{}}
\toprule
Series & 2006 & 2007 & 2008 & 2009 & 2010 & 2011 & 2012 & 2013 & 2014 & 2015 & Mean \\ 
\midrule
\multicolumn{6}{l}{\hspace{-.13in} Japan (RMSFE)} \\
Female & 0.93 & 0.95 & 1.26 & 0.89 & 1.49 & 1.55 & 1.43 & 1.30 & 1.03 & 1.20 & 1.20\\ 
  Male & 0.95 & 1.07 & 2.01 & 0.98 & 2.10 & 1.81 & 1.71 & 1.03 & 0.80 & 1.20 & 1.37\\ 
\\
\multicolumn{6}{l}{\hspace{-.13in} Japan (Mean interval score)} \\
Female & 1.81 & 2.13 & 3.40 & 1.75 & 4.99 & 4.55 & 3.55 & 2.80 & 1.70 & 2.73 & 2.94\\ 
 Male & 2.39 & 3.05 & 6.73 & 2.61 & 7.87 & 6.18 & 5.61 & 2.59 & 1.89 & 3.40 & 4.23\\ 
\\
  \multicolumn{6}{l}{\hspace{-.13in} Japanese prefectures (RMSFE)} \\
Female & 1.28 & 1.28 & 1.44 & 1.21 & 1.63 & 1.69 & 1.76 & 1.50 & 1.23 & 1.89 & 1.49 \\ 
  Male & 2.72 & 2.71 & 3.00 & 2.73 & 3.26 & 3.20 & 2.93 & 2.60 & 2.20 & 2.22 & 2.76\\ 
\\
\multicolumn{6}{l}{\hspace{-.13in} Japanese prefectures (Mean interval score)} \\
Female & 2.79 & 2.71 & 3.35 & 2.64 & 3.94 & 4.12 & 4.34 & 3.17 & 3.07 & 3.05 & 3.32\\ 
Male & 7.18 & 6.74 & 7.94 & 6.92 & 8.79 & 8.58 & 7.73 & 6.18 & 5.42 & 5.63 & 7.11\\ 
  \bottomrule
\end{tabular}
\end{table}

Compared to our proposal model averaging method, the existing model averaging method assigns different weights to different models. From the in-sample forecast errors, the existing model averaging method assigns higher weights for those more accurate models and lower weights for those less accurate models. By contrast, our proposed model averaging method selects a superior subset of models and assign equal weights. From the results, we find that the proposed model averaging method gives a smaller forecast error than the existing model averaging method.

\section{Discussion}\label{sec:7}

We first revisit four families of stochastic mortality models, namely the Renshaw-Haberman models, the Cairns-Blake-Dowd models, the Lee-Carter models, and the functional time-series models. From the viewpoint of actuarial science, mortality forecasts are an important input for determining annuity prices and reserves. From the viewpoint of demography, mortality forecasts are vital for policy-making at the national and sub-national levels. Using the national and sub-national Japanese mortality rates, we evaluate and compare point and interval forecast accuracies, as measured by the root mean squared error and mean interval score, among the 17 time-series extrapolation methods and two model-averaging methods considered.

From the viewpoint of the point forecast accuracy, the Plat model gives the smallest point forecast errors, followed by the Lee-Carter and model-averaged method for Japanese females and males. In this case, because the superior set of the models was determined from in-sample forecast errors, it is the case where the best model for the in-sample forecasts may not be the best model for out-of-sample forecasts. This affects the forecast accuracy of the model-averaged method. Also, the Lee-Carter method with the Poisson error structure is more accurate than the version with the Gaussian error structure. For modeling sub-national Japanese females, the Plat model also performs the best; while the multilevel functional time-series model performs the best for males. 

From the viewpoint of the interval forecast accuracy, the Plat model gives the smallest interval forecast errors, followed by the model-averaged methods for Japanese females. For Japanese males, the model-averaged methods produce the smallest interval forecast errors for Japan and are on a par with the product-ratio method for Japanese sub-national data. To our surprise, the Renshaw-Haberman methods produce relatively worse interval forecast accuracy than that produced by the Lee-Carter and functional time-series methods. The result could be due to the instability of parameter estimation. 

From Tables~\ref{tab:24},~\ref{tab:24b} and~\ref{tab:240},~\ref{tab:240b}, the best model for producing point forecasts does not necessary the same as the best model for producing interval forecasts, and the best model for producing the national series does not necessarily perform the best for the sub-national series, as the features of the data may be different.

Because different models have their advantages and disadvantages, we apply a model-averaged method to select a set of superior model based on the model confidence set. The model confidence set is a procedure that determines a set of superior models based on the in-sample forecast errors.

For producing point forecasts, the selected superior sets of models are more diverse. For producing interval forecasts, the selected superior set of models often includes the product-ratio method, especially for male series. Given the product-ratio method is a joint modeling and coherent forecasting method, it achieves better forecast accuracy for males with a small sacrifice in terms of the female results. By coherent, we believe within each prefecture, female and male subpopulations share the similar characteristics, such as health facilities. Generally, the joint modeling methods, which also include the multivariate and multilevel functional time-series methods, often but not always outperform the model without incorporating correlation among subpopulations. The model-averaged forecasts may not perform the best for females and males, but they tend to give an aggregate best performance.

We show that potential gains in forecast accuracy can be achieved by discarding the worse performing models before combining the forecasts equally. By contrast, an existing model-averaging method assigning different weights for all models performs worse than the proposed method. We find that the proposed model-averaging method offers a more robust procedure for selecting the forecasting models based on their in-sample performances. By robustness, the model-averaged methods are protected against model misspecification. The advantage of the model-averaged methods is more apparent for males than females.

The accurate forecasting of mortality at retirement ages is essential to determine life, fixed-term and delayed annuity prices for various maturities and starting ages \citep[see, e.g.,][]{SH17}. In the online supplement, we present a study on calculating single-premium fixed-term immediate annuity. To forecast mortality rates, we suggest considering the notion of model averaging.

In our modeling and analysis, we have made several choices. Below, we set out the ways in which the different choices could potentially have affected our results/overall conclusions.
\begin{enumerate}
\item[1)] We considered ages from 60 and 100+ to study the mortality pattern of retirees. We could apply the model averaging idea to other age groups, such as ages from 0 to 100+. With different age groups, the point and interval forecast results may be different. 
\item[2)] Other age-specific mortality forecasting models could be incorporated into the model averaging. We consider only time-series extrapolation methods, and did not consider the expectation or explanation methods. For long-term forecasts and the expectation, expectation method has been used by the \cite{CMI18}. The inclusion of the expectation or explanation methods may alter the selection of superior models.
\item[3)] We modeled age-specific mortality rates, but the focus could also be on death counts or survival probabilities. For example, when we model age-specific death counts, there are other models, such as compositional data analysis, that could be included in the initial model pool.
\item[4)] After selecting the superior set of models, one could assign different weights instead of equal weights considered. With different weights, our forecast results may be improved as considered in \cite{Shang12}.
\item[5)] In applying the model confidence set procedure of \cite{HLN11}, we consider the 90\% confidence level. Considering other levels of confidence is possible. In general, as the confidence level increases, the number of superior models decreases.
\item[6)] We evaluated and compared one-step-ahead, five-step-ahead, and ten-step-ahead point and interval forecast accuracies. Considering other forecast horizons is also possible. For the longer term, the extrapolation methods may not perform well. 
\item[7)] We evaluated point forecast accuracy by the root mean square error and interval forecast accuracy by the mean interval score and coverage probability deviance, respectively. Considering other forecast error criteria, such as mean absolute percentage error or mean absolute scaled error, is also possible. With different error measures, the point and interval forecast results may be different.
\end{enumerate}

We believe the present work paves the way for the above possible future research directions, and the proposed model-averaging method should be a welcome addition to the demographic modeling and forecasting.

\newpage
\begin{center}
{\large\bf SUPPLEMENTARY MATERIAL}
\end{center}

\begin{description}

\item[Code for Shiny application] The R code to produce a Shiny user interface for plotting every series in the Japanese human mortality data. (shiny.R)

\item[Geography locations of the 47 prefectures in Japan] We present a graphical display of the 47 prefectures within eight regions in Japan and include a table documenting the names of the 47 prefectures. (supplement\_MCS.pdf)

\item[Detailed point and interval forecast results] While Table~\ref{tab:4} presents a summary of the point and interval forecast accuracies, we present the detailed forecast results for ten years in the forecasting period. (supplement\_MCS.pdf)

\item[Calculation for single-premium fixed-term immediate annuity] The forecasted mortality rate is an essential input for determining temporary annuity prices for various maturities and starting ages of the annuitant. (supplement\_MCS.pdf) 

\end{description}

\newpage
\bibliographystyle{agsm}
\bibliography{MCS}

\end{document}